# Thermodynamic Analyses of the LCLS-II Cryogenic Distribution System

Andrew Dalesandro, Joshua Kaluzny, Arkadiy Klebaner, *Fermi National Accelerator Laboratory*

***Abstract*— The Linac Coherent Light Source (LCLS) at Stanford Linear Accelerator Center (SLAC) is in the process of being upgraded to a superconducting radio frequency (SRF) accelerator and renamed LCLS-II. This upgrade requires thirty-five 1.3 GHz SRF cryomodules (CM) and two 3.9 GHz CM. A cryogenic distribution system (CDS) is in development by Fermi National Accelerator Laboratory to interconnect the CM Linac with the cryogenic plant (CP). The CDS design utilizes cryogenic helium to support the CM operations with a high temperature thermal shield around 55 K, a low temperature thermal intercepts around 5 K, and a SRF cavity liquid helium supply and sub-atmospheric vapor return both around 2 K. Additionally the design must accommodate a Linac consisting of two parallel cryogenic strings, supported by two independent CP utilizing CDS components such as distribution boxes, transfer lines, feed caps and endcaps. The paper describes the overall layout of the cryogenic distribution system and the major thermodynamic factors which influence the CDS design including heat loads, pressure drops, temperature profiles, and pressure relieving requirements. In addition the paper describes how the models are created to perform the analyses.**

***Index Terms*— Cryogenic fluids, Fluid flow, Liquid helium, Pressure effects, Superconducting linear accelerators**

## I. Introduction

THE LINAC Coherent Light Source II (LCLS-II) located at SLAC National Accelerator Laboratory (SLAC) in Menlo Park, CA is a U.S. Department of Energy project tasked to design and build a world-class x-ray free-electron laser facility for scientific research. The LCLS-II accelerator (Linac) design is based on superconducting radio frequency (SRF) technology employing thirty-five 1.3 GHz SRF cryomodules and two 3.9 GHz SRF cryomodules in continuous wave operation. The LCSL-II cryogenic system consists of three major subsystems: cryogenic plant (CP), cryomodules (CM), and cryogenic distribution system (CDS). The CDS supplies cryogens from the CP to CM Linac, with interfaces to both systems. The CDS design presented herein is based on a reference baseline design, which may vary marginally from the final delivery. The CDS includes the following subcomponents:

- Two – Distribution boxes (DB-U, DB-D), distribute flow and provide low temperature heat exchanger
- Six – Feed caps (FC-1, FC-2, FC-3, FC-4, FC-5, FC-6), connect CM primary cryogenic circuits to CDS transfer lines
- Two – End caps (EC-U, EC-D), return cryogens at the end of the Linac
- Two – Cryogenic bypass transfer lines (LH, BC1), allow tunnel floor space for warm beam line equipment
- Two – Vertical transfer lines (VTL-U, VTL-D), connect the CDS in the tunnel to the DB upstairs
- Two – Surface transfer lines (STL-U, STL-D), connect the DB to the CP

The cryogenic system has an upstream and downstream section, where upstream represents the source of the LCLS-II electron beam. Each stream has a dedicated CP, DB, and CM Linac (upstream: L0, L1, L2; downstream: L3). The upstream Linac has a total of 15 CM while the downstream Linac has a total of 20 CM. Fig. 1 provides an upper level overview of the LCLS-II cryogenic system upstream and downstream, respectively, including the CDS.

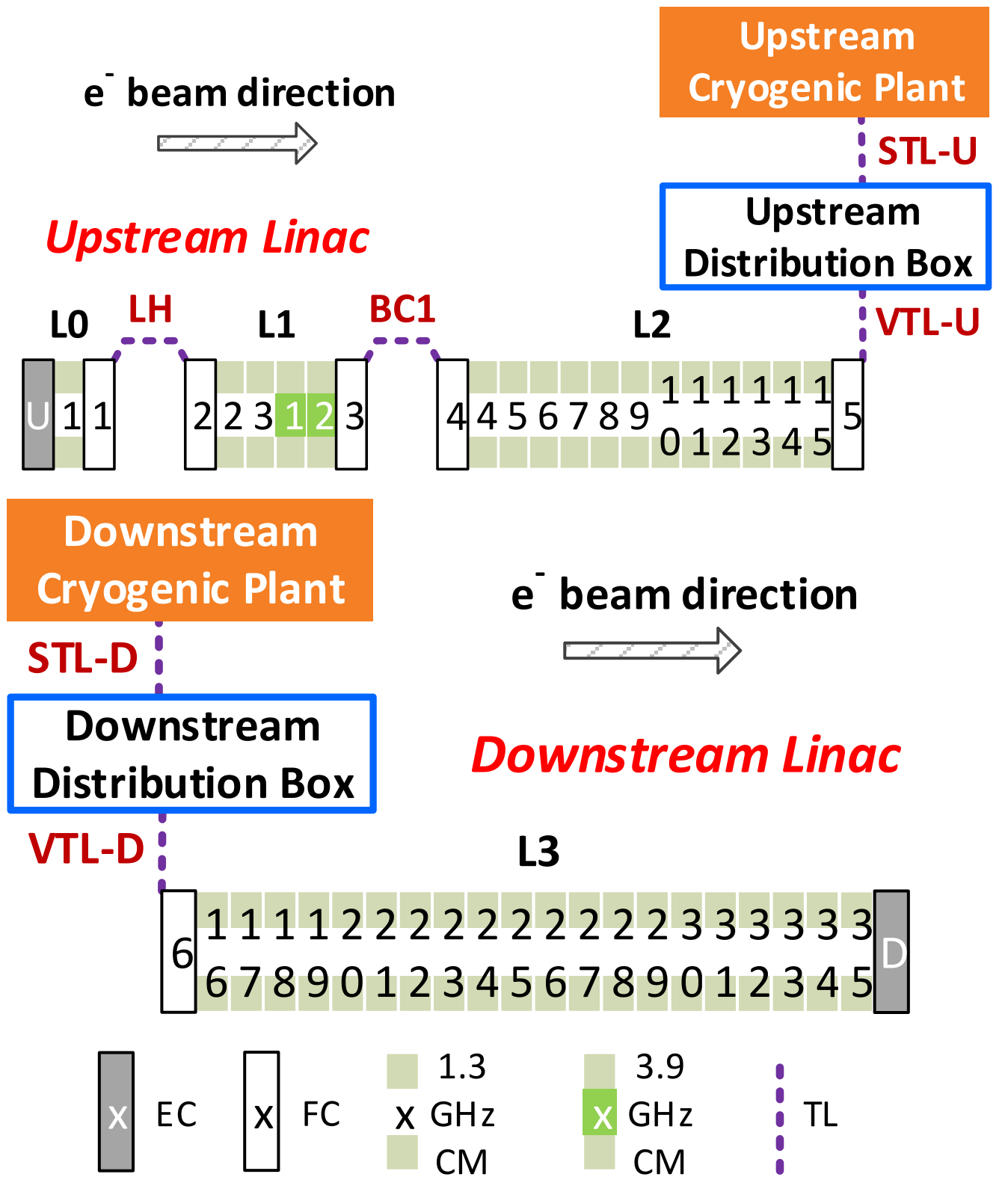

Fig. 1. LCLS-II cryogenic system overview schematic, including CDS and CM Linacs, segmented by upstream and downstream strings.

A. Dalesandro is with the Fermi National Accelerator Laboratory, Batavia, IL 80510 USA (e-mail: andrewd@fnal.gov).
J. Kaluzny is with the Fermi National Accelerator Laboratory, Batavia, IL 80510 USA (e-mail: kaluzny@fnal.gov).
A. Klebaner is with the Fermi National Accelerator Laboratory, Batavia, IL 80510 USA (e-mail: klebaner@fnal.gov).

## II. Operating Requirements

With two distinct CM Linac strings operating simultaneously and identical cryogenic operating requirements for each, the CDS requires flexibility to accommodate differences in the Linac geometries and flow requirements. All CDS subcomponents contain six primary cryogenic process circuits, named alphabetically, Lines A-F, which correspond to the primary cryogenic circuits of the CM Linac, as described below. Operating parameters of each CDS Linac at the interface between the CDS and CP are provided in Table 1 [1].

- Line A – 4.5 K subcooled helium cavity supply
- Line B – 2 K subatmospheric cavity helium return
- Line C – 5 K low temperature helium intercept supply
- Line D – 8 K low temperature helium intercept return
- Line E – 35 K high temperature shield helium supply
- Line F – 55 K high temperature shield helium return

Table 1
CDS Process Circuit Operating Parameters at CDS/CP Interface

| | Upstream | | | Downstream | | |
|---|---|---|---|---|---|---|
| Circuit | Temperature | Pressure | Flow Rate | Temperature | Pressure | Flow Rate |
| - | K | bar | g/s | K | bar | g/s |
| Line A | 4.5 | 3.2 | 90 | 4.5 | 3.2 | 98 |
| Line B | 3.4 | 2.9E-02 | 90 | 3.4 | 2.9E-02 | 98 |
| Line C | 5.5 | 3.2 | 26 | 5.5 | 3.2 | 27 |
| Line D | 7.5 | 3.2 | 26 | 7.5 | 3.2 | 27 |
| Line E | 35 | 3.7 | 78 | 35 | 3.7 | 66 |
| Line F | 55 | 3.3 | 78 | 55 | 3.5 | 66 |

The CDS is designed as an integrated part of the cryogenic system with capacity parameters and constraints defined by both CP and CM Linac. The primary operating constraints on the CDS design are the pressure drop and heat leak budgets, with each presented in Table 2 and provided by LCLS-II management [2].

Table 2
CDS Process Circuit Budgets - Pressure Drop and Heat Leak

| Circuit | Temperature | Pressure Drop Budget | Heat Load Budget |
|---|---|---|---|
| - | K | bar | W |
| Line A | 2 | 1.0 | 290 |
| Line B | | 4.0E-03 | |
| Line C | 5 | 1.0 | 280 |
| Line D | | | |
| Line E | 35 | 1.5 | 4,350 |
| Line F | | | |

## III. Thermodynamic Analysis

### A. Pressure Drop

The CDS thermodynamic design is based on a piping model of the pressure drop of each process circuit that includes all CDS and CM Linac subcomponents with flow parameters updated at each piping element. The basic pressure drop equation (1) is taken from Crane [3] with helium fluid properties (density: $\rho$) calculated within the model using HEPAK [4]. Other inputs include mass flow: $\dot{m}$, friction factor: $f$, pipe diameter: $d$, pipe length: $L$.

$$\Delta P = 3.36\text{x}10^{-6} + \frac{fL(\dot{m} \text{ x } 7.937)^2}{d^5\rho} \tag{1}$$

The LCLS-II CP utilizes cold compression to achieve efficient refrigeration at 2 K at each CM cavity. Any reduction in pressure at compressor suction reduces the operating capacity. It is for this reason that the pressure drop budget for Line B is only 4 mbar, and as such the majority of the pressure drop analysis is focused on Line B. The CDS Line B pressure drop model is based on initial conditions presented below at each End Cap [5]. The mass flow is increased uniformly at each CM, starting with the CM nearest the adjacent EC, until the operating flow rate for a given Linac, as presented in Table 1, is achieved at the final CM nearest the DB. Fig. 2 and (2)-(8) provide a simplified overview of how the mass and energy balance is handled at each CM. Initial and final properties are denoted with subscripts $i$ and $f$, respectively. Initial conditions are assumed at each CDS EC include temperature, $T_i$ = 2.0 K, pressure, $P_i$ = 31 mbar, enthalpy, $h_i$ = 25 J/g, and mass flow, $m_i$ = 0 g/s. Other inputs include loss coefficient: K, and fluid velocity: v.

$$\dot{m}_{add} = \dot{m}_Q + \dot{m}_{JT} \tag{2}$$

$$\dot{Q}_{LHe} = \sum \dot{Q}_{static|cavities} + \dot{Q}_{cavity\ RF} \tag{3}$$

$$\dot{m}_Q = \frac{\dot{Q}_{LHe}}{h_{fg}} \quad \text{where} \quad h_{fg} = \text{He latent heat} \tag{4}$$

$$\dot{m}_{JT} = \frac{x_S * \dot{m}_Q}{1 - x_S} \quad \text{where} \quad x_s = \frac{\dot{m}_{JT}}{\dot{m}_{add}} \tag{5}$$

$$\dot{Q}_{add} = \sum \dot{Q}_{static|CDS\ components} \tag{6}$$

$$h_f = \frac{\dot{m}_i h_i + \dot{m}_{add} h_{vap} + \dot{Q}_{add}}{\dot{m}_f} \tag{7}$$

$$\Delta P = \frac{8\dot{m}_f{}^2 fL}{\pi^2 \rho D^5} + \frac{\rho K v_f{}^2}{2} + \frac{\rho (v_f{}^2 - v_i{}^2)}{2} \tag{8}$$

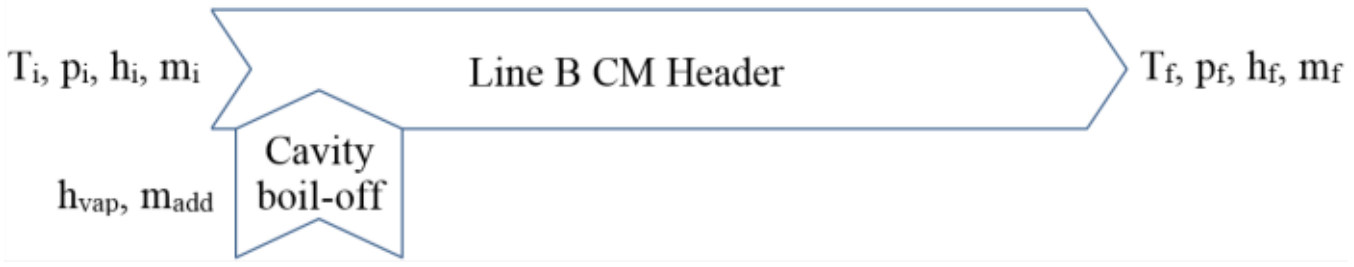

Fig. 2. Line B Simplified Mass Flow Concept [5]

### B. Heat Load

The CDS pressure drop model applies heat load at each component based on a combination of published loads for comparable systems [6]-[9] and calculations resulting from the CDS system geometry. Fig. 3 provides a basic heat load schematic applied in the model to estimate helium boil-off rate and temperature within the CM Lines A and B, including the heat exchanger balance. Other inputs to the baseline heat load include heat exchanger effectiveness, $\varepsilon$, of 90%, SRF cavity quality factor, $Q_0$, of $2.7\text{x}10^{10}$, and accelerating gradient, $E$, of 16 MV/m.

### C. Pressure Safety

The CDS provides pressure safety of all CDS components as well as the cryogenic process circuits of each CM Linac. The approach to pressure safety is based on the assumed worst case single failure scenario for a given system, including overfill from the CP, overflow through the CM supply valves, loss of beam vacuum, and loss of insulating vacuum [10]. Loss of vacuum heat flux values are specified by the LCLS-II project [11]. Credit is taken for segmented insulating vacuum and for the propagation delay experienced by subsonic heat waves during loss of vacuum in comparable geometries due to cryopumping. Pressure safety design requirements are presented in Table 3.

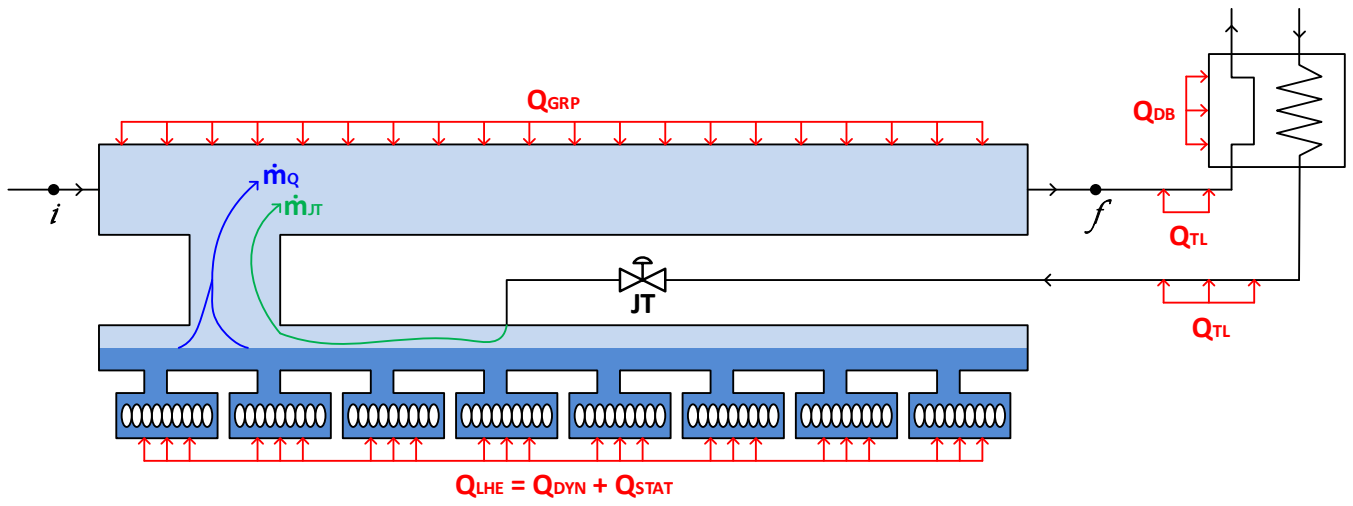

Fig. 3. Single CM Simplified Heat Load Schematic [5]

TABLE 3
PRESSURE SAFETY REQUIREMENTS

| Circuit | Location | Type | Set Pressure | Upstream Pressure | Allowable Overpressure | Relief Temperature | Required Vent Rate |
|---|---|---|---|---|---|---|---|
| - | - | - | bar | bar | bar | K | g/s |
| Line A | DB-U | SV | 20 | 20.5 | 24 | 12.7 | 568 |
| | DB-D | SV | 20 | 20.2 | 24 | 12.7 | 481 |
| Line B | DB-U | SV | 2.05 | 2.10 | 2.16 | 80 | 645 |
| | | RD | 4.1 | 4.2 | 4.8 | 6.9 | 7,907 |
| | DB-D | SV | 2.05 | 2.08 | 2.16 | 80 | 645 |
| | | RD | 4.1 | 4.2 | 4.8 | 6.9 | 7,580 |
| Line C, D | DB-U | SV | 20 | 20.6 | 24 | 12.6 | 576 |
| | DB-D | SV | 20 | 20.2 | 24 | 12.7 | 399 |
| Line E, F | DB-U | SV | 13 | 13.2 | 24 | 55 | 146 |
| | DB-D | SV | 13 | 13.1 | 24 | 55 | 146 |

## IV. RESULTS

The CDS analysis results herein present the pressure drop, pressure profile, mass flow, and temperature profiles at the baseline operating parameters for the LCLS-II cryogenic system as discussed in this paper, unless otherwise specified. Fig. 4 and Fig. 5 present the pressure drop for the 2 K cold compressor suction header Line B vs. the cavity quality factor (which is an indicator of heat load), heat exchanger effectiveness, and beam accelerating gradient (which is an indicator of RF power) for US and DS Linacs, respectively. Fig. 6 and Fig. 7 present the US and DS Linacs, respectively, 2 K circuit mass flow rate vs. similar cavity and heat exchanger parameters as Fig. 4 and Fig. 5. Fig. 8 presents the Line A aggregate pressure drop and Fig. 9 presents the Line B pressure profile, both vs. the Linac position and both include the mass flow rate. Fig. 10 and Fig. 11 present the accumulated pressure drop, temperature profile, and mass flow rate for the 5 K circuit (Line C and D) and the HTS circuit (Line E and F) vs. Linac position, respectively. Fig. 10 and Fig. 11 each have two points along the Linac because the 5 K and HTS circuits each have a supply and return line. Fig. 12 presents the temperature profile and mass flow of the 2 K system at each CM and CDS subcomponent.

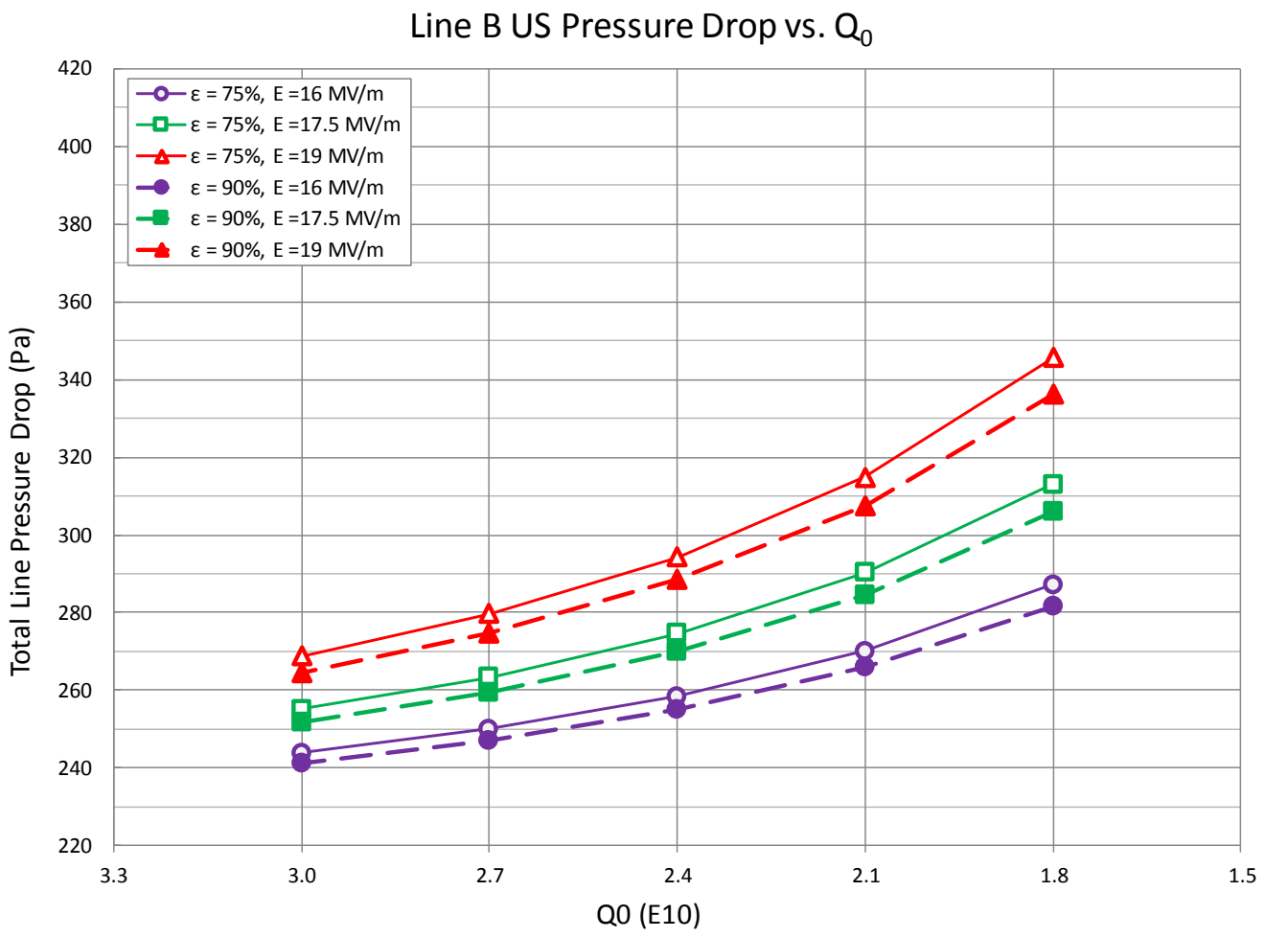

Fig. 4. Line B Upstream Pressure Drop vs. Cavity Quality Factor, *Q0*, with varying heat exchanger effectiveness, *ε*, and accelerating gradient, *E*.

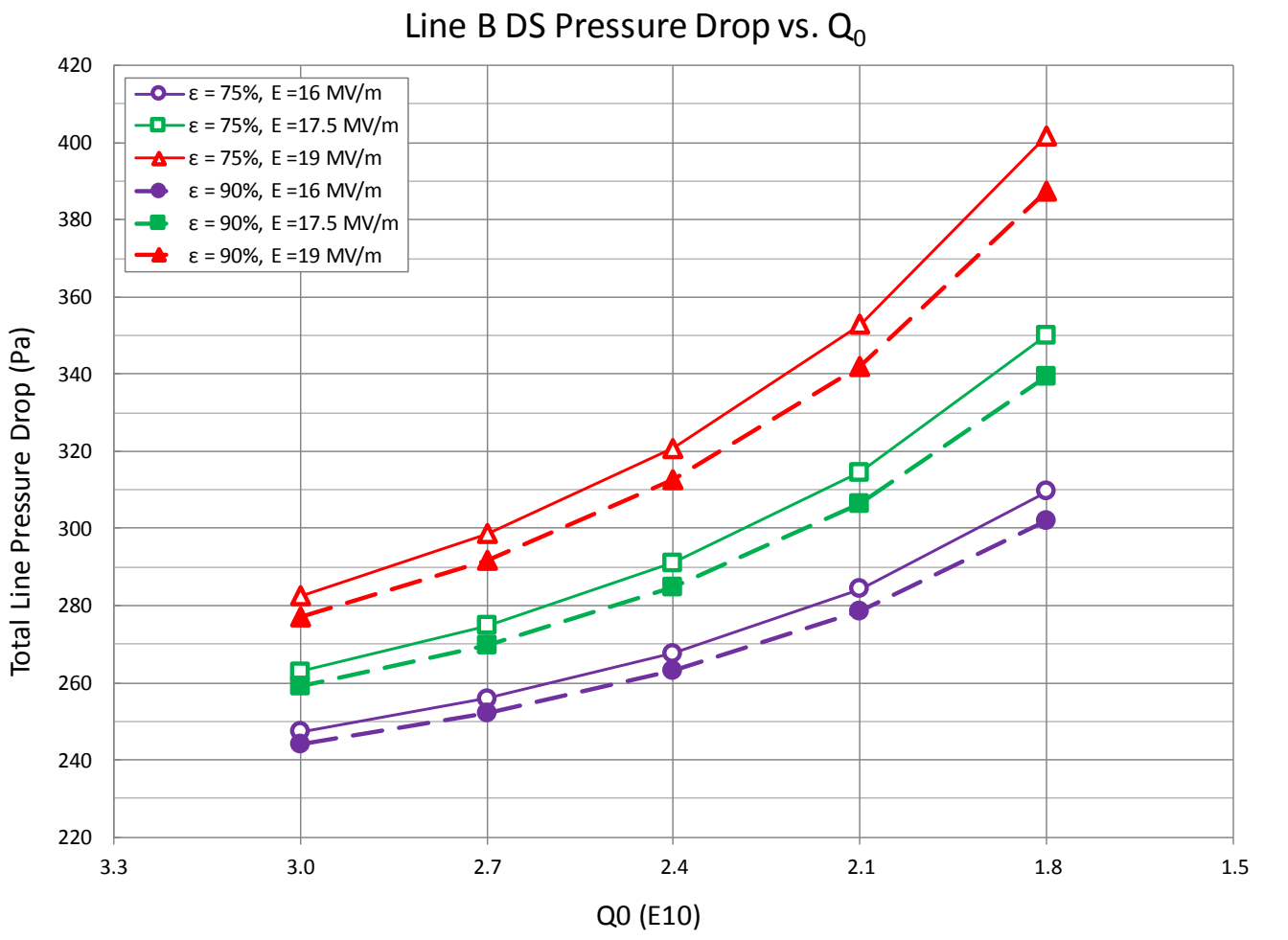

Fig. 5. Line B Downstream Pressure Drop vs. Cavity Quality Factor, *Q0*, with varying heat exchanger effectiveness, *ε*, and accelerating gradient, *E*.

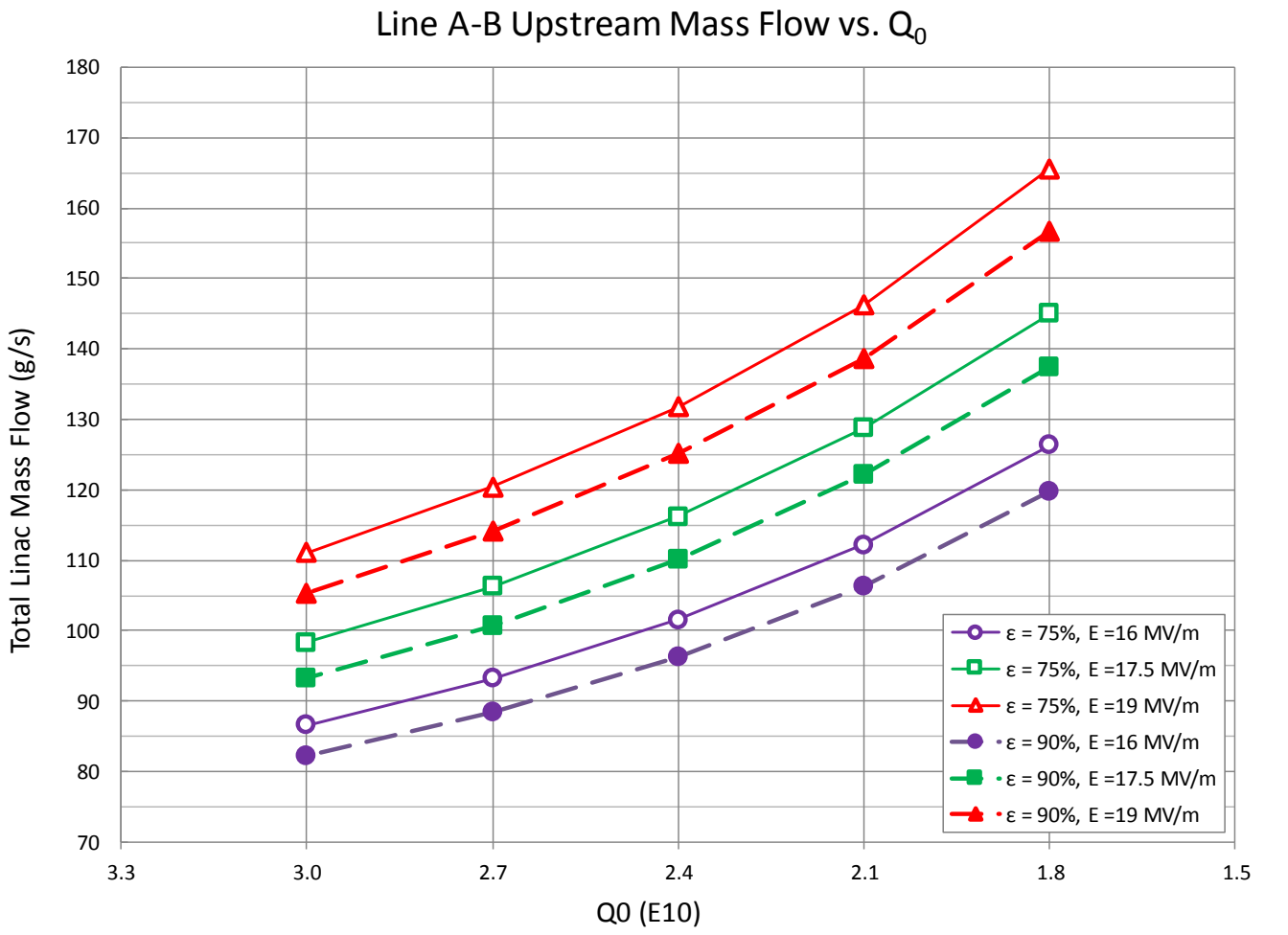

Fig. 6. 2 K Circuit Upstream Mass Flow vs. Cavity Quality Factor, *Q0*, with varying heat exchanger effectiveness, *ε*, and accelerating gradient, *E*. [12]

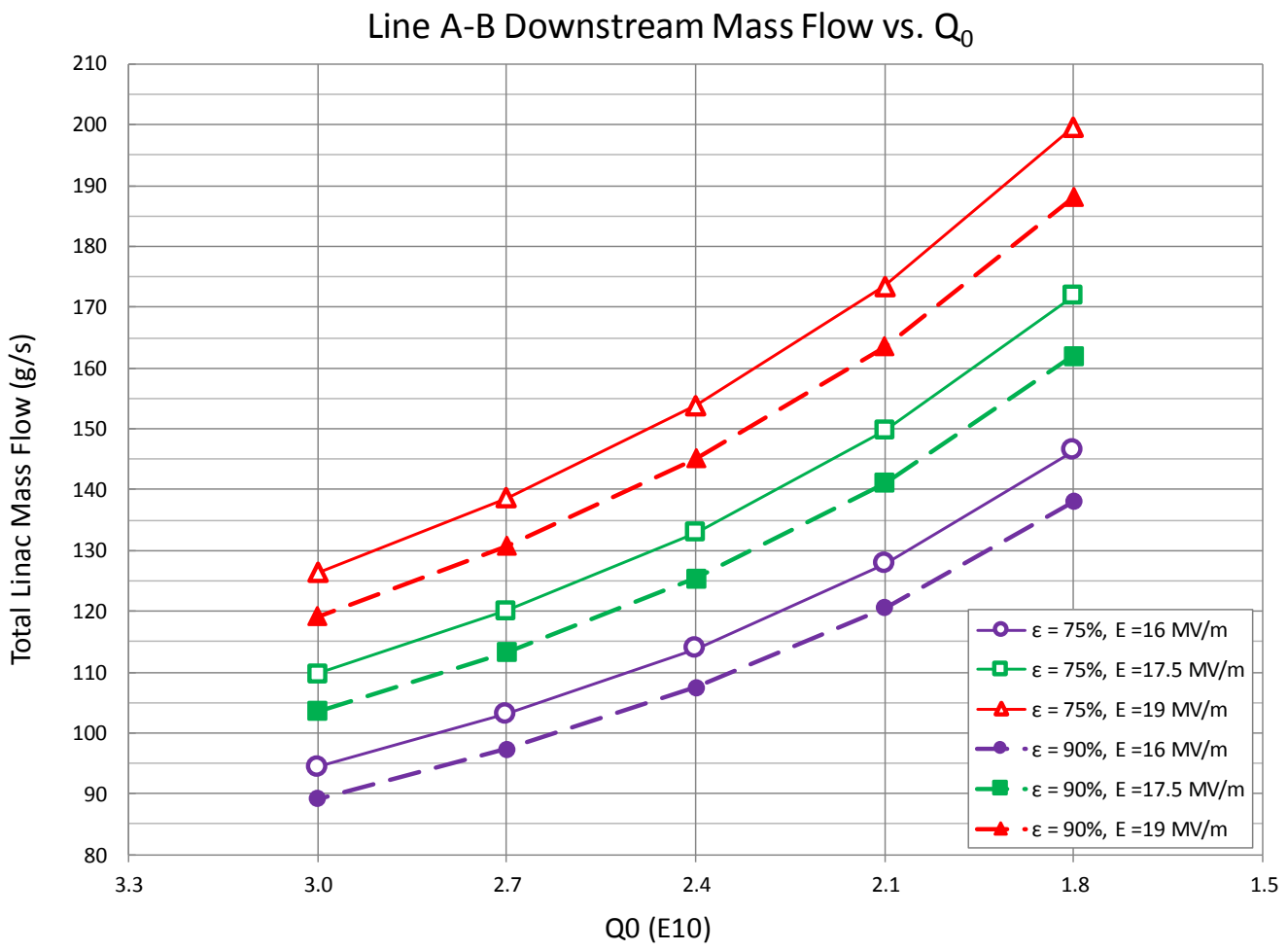

Fig. 7. 2 K Circuit Downstream Mass Flow vs. Cavity Quality Factor, *Q0*, with varying heat exchanger effectiveness, *ε*, and accelerating gradient, *E*. [12]

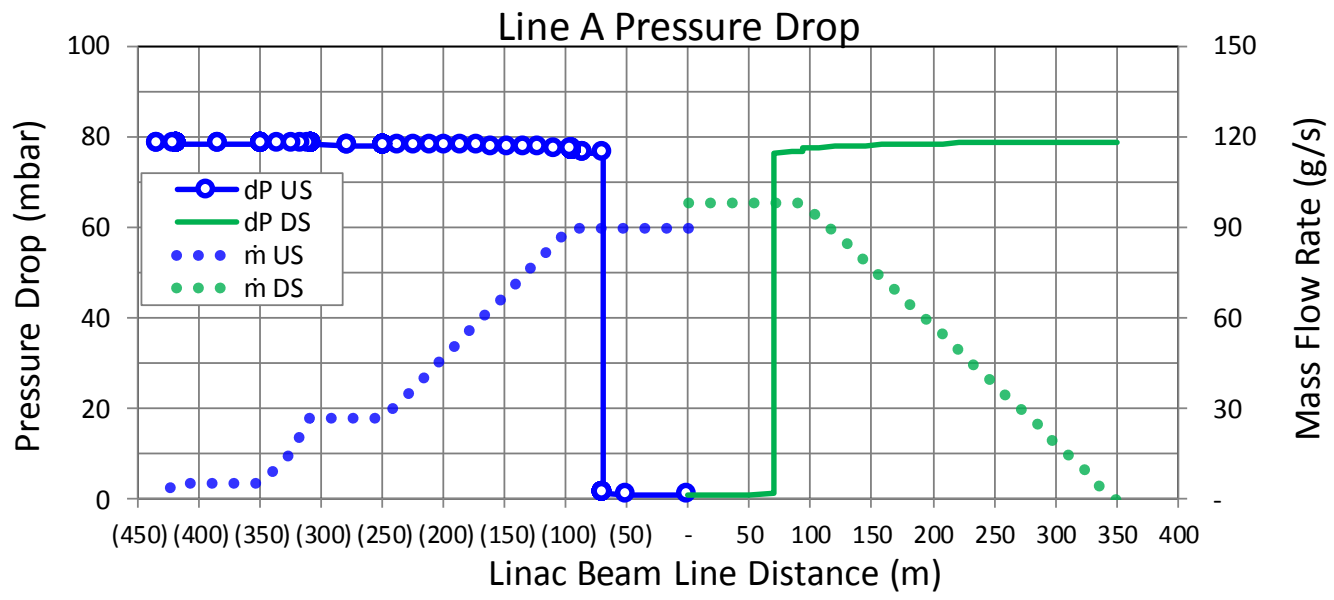

Fig. 8. Line A Aggregate Pressure Drop and Mass Flow for CDS Linacs Upstream (US), left, and Downstream (DS), right, vs. Linac distance. Linac position 0 corresponds to the midpoint of the Linac, at the CDS DB. [1]

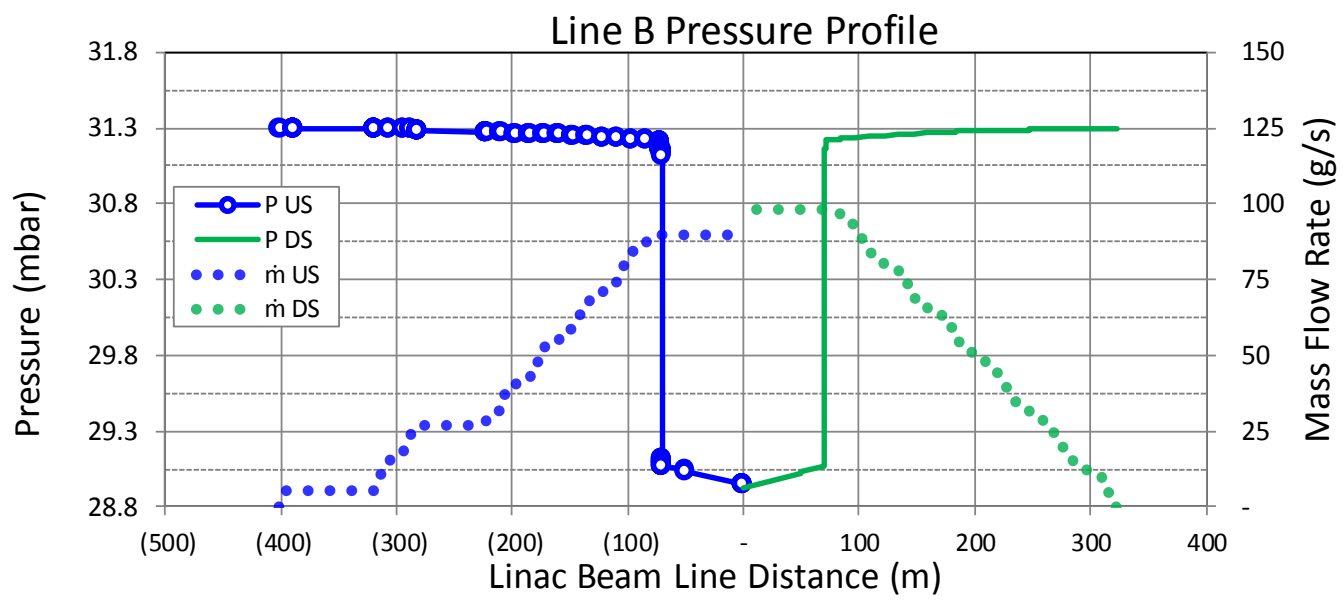

Fig. 9. Line B Pressure Profile and Mass Flow for CDS Linacs Upstream (US), left, and Downstream (DS), right, vs. Linac distance. Linac position 0 corresponds to the midpoint of the Linac, at the CDS DB. [5]

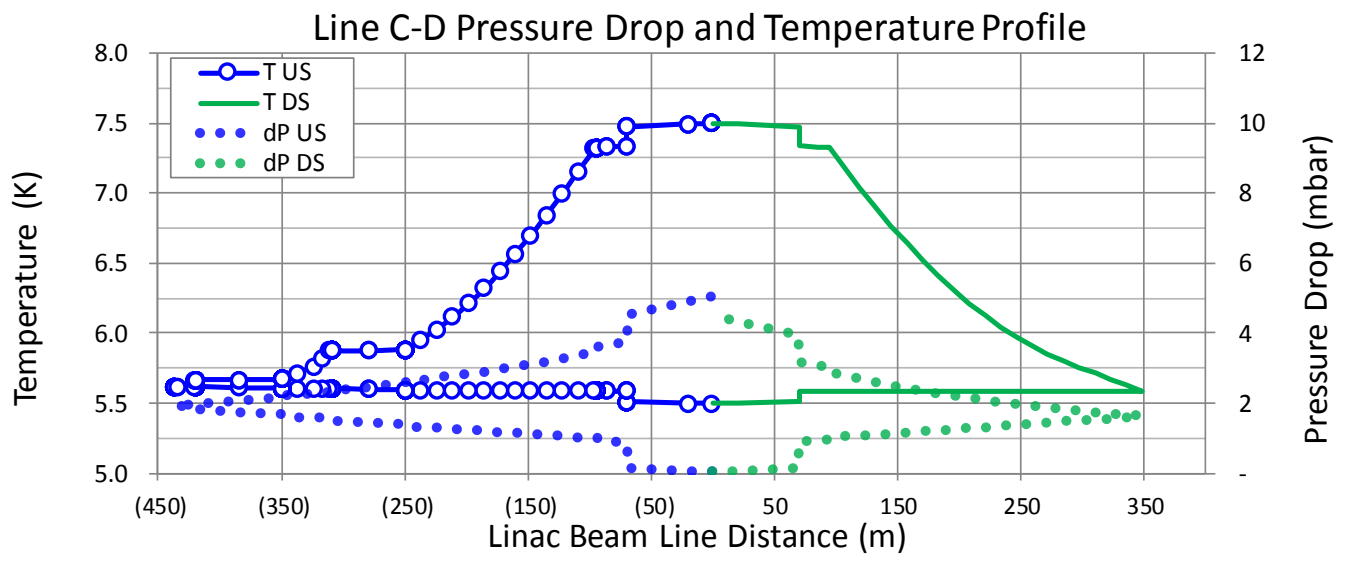

Fig. 10. Line C and D Combined Pressure and Temperature Profiles for CDS Linacs Upstream (US), left, and Downstream (DS), right, vs. Linac distance. Linac position 0 corresponds to the midpoint of the Linac, at the CDS DB. [1]

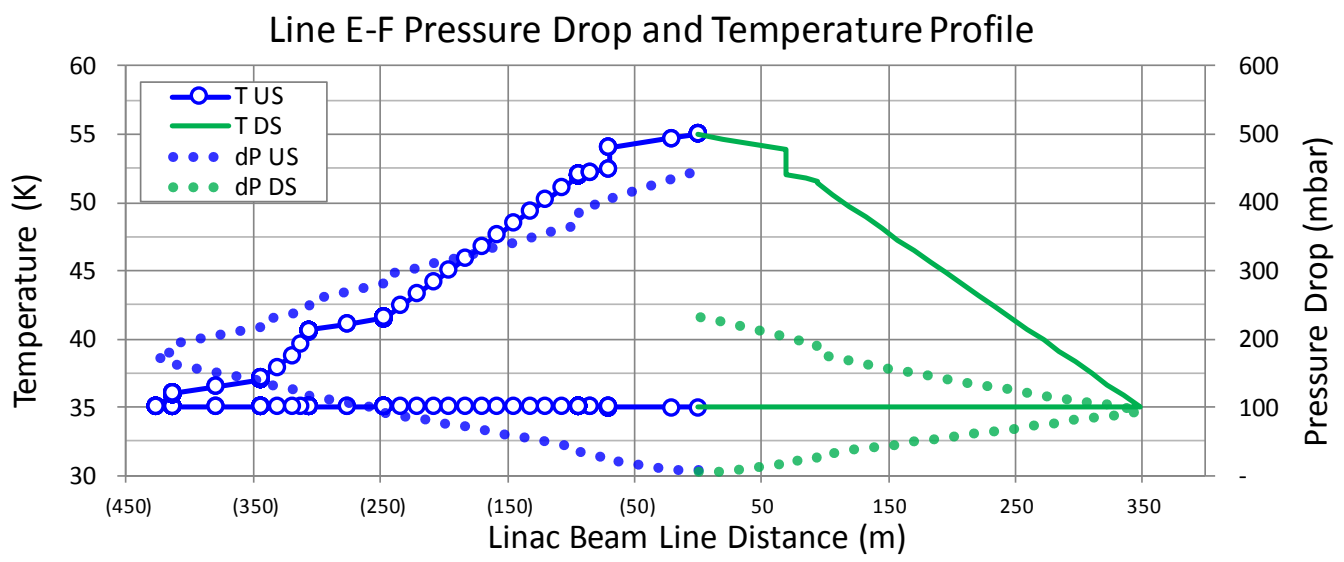

Fig. 11. Line E and F Combined Pressure and Temperature Profile for CDS Linacs Upstream (US), left, and Downstream (DS), right, vs. Linac distance. Linac position 0 corresponds to the midpoint of the Linac, at the CDS DB. [1]

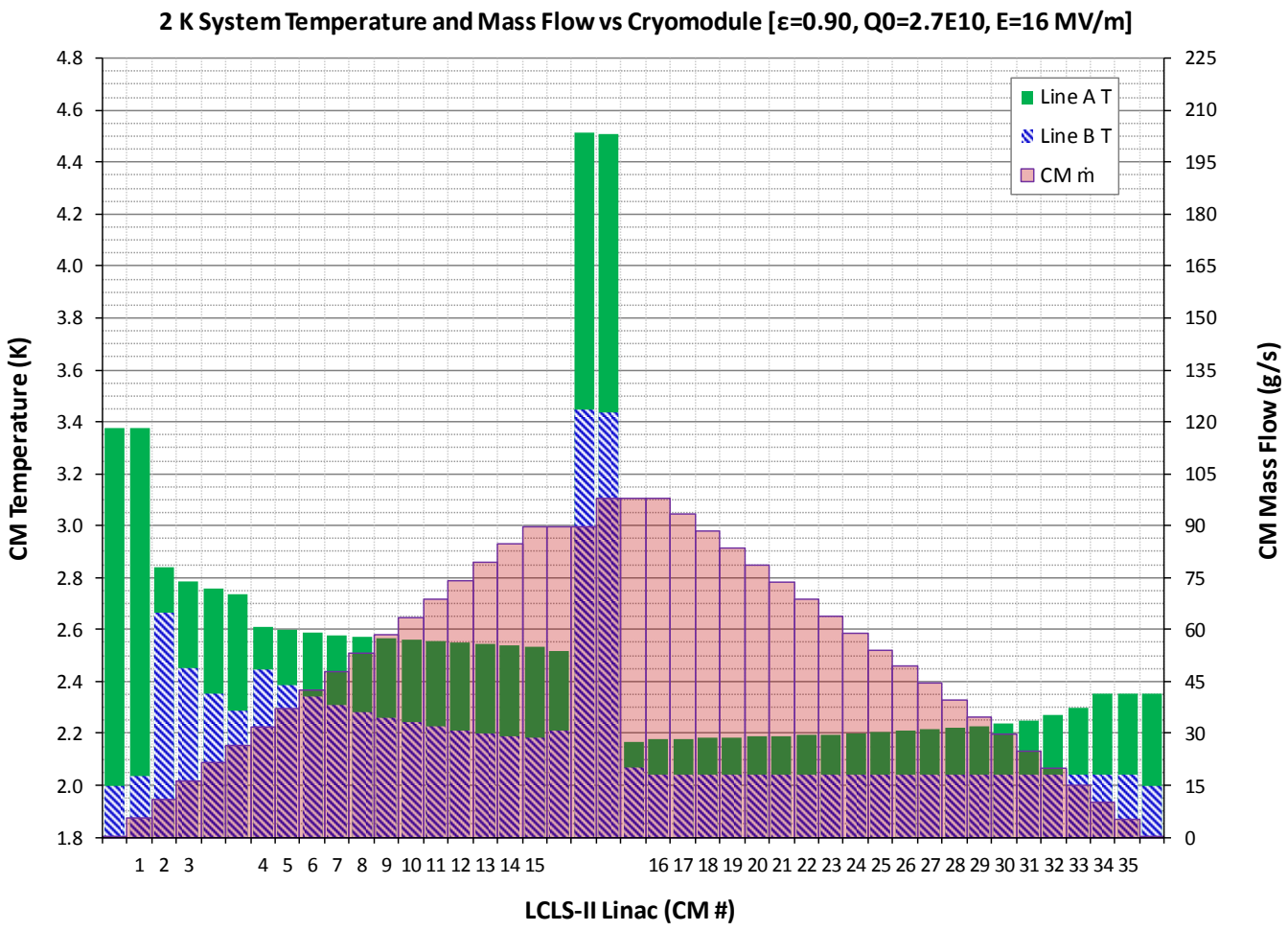

Fig. 12. 2 K Circuit Temperature and Mass Flow Throughout CDS vs. Each LCLS-II CM. Line A temperature exceeds Line B temperature at all points. [12]

## V. Conclusion

The CDS is designed with sufficient margin for all LCSL-II cryogenic system performance and operating budgets while considering a variety of CM, heat exchanger, and heat load scenarios beyond of the stated project baseline operating parameters [2]. The CDS is sufficiently robust to accommodate all cryogenic system operating and transient design modes.

## References

[1] "LCLS-II CDS HT Shield, LT Intercept and 2K Supply dP Analysis," LCLSII-4.9-EN-0295, 2015.

[2] "Cryogenic Distribution System Functional Requirements Specification," LCLSII-4.9-FR-0057, 2014.

[3] Crane, "Flow of Fluids Through Valves, Fittings, and Pipe," New York, New York: Eighteenth Printing, 1979.

[4] Horizon Technologies, HEPAK v3.4, 2011.

[5] "LCLS-II CDS Gas Return Pipe Pressure Drop Analysis," LCLSII-4.9-EN-0292, 2015.

[6] "LCLS-II Cryogenic Heat Load," LCLS-II-4.5-EN-0179, 2014.

[7] J. G. Weisand, "TESLA Cryomodule Operating Experience and Design Choices," Fermilab, 2001.

[8] T. H. Nicol, "TESLA Test Cell Cryostat Support Post Thermal and Structural Analysis," FERMILAB-TM-1794, Fermilab, 1992.

[9] "LCLS-II CDS Heat Leak Analysis," LCLSII-4.9-EN-0299, 2015.

[10] "LCLS-II CDS Relief System Analysis," LCLSII-4.9-EN-0300, 2016.

[11] "LCLS-II Relieving Heat Flux Requirements," LCLSII-4.5-EN-0214, 2014.

[12] "LCLS-II CDS 2 K System and HX Parameters Analysis", LCLSII-4.9-EN-0516, 2015.